\documentclass[letter,twocolumn]{jpsj3arXiv}
\usepackage{txfonts}

\title{Lock-in of a Chiral Soliton Lattice by Itinerant Electrons}

\author{Shun Okumura\thanks{s.okumura@aion.t.u-tokyo.ac.jp}, Yasuyuki Kato, and Yukitoshi Motome}
\inst{Department of Applied Physics, The University of Tokyo, Hongo, Bunkyo, Tokyo 113-8656, Japan} 

\abst{
Chiral magnets often show intriguing magnetic and transport properties associated with their peculiar spin textures.
A typical example is a chiral soliton lattice, which is found in monoaxial chiral magnets, such as ${\rm CrNb_3S_6}$ and ${\rm Yb(Ni_{1-{\it x}}Cu_{{\it x}})_3Al_9}$ in an external magnetic field perpendicular to the chiral axis.
Here, we theoretically investigate the electronic and magnetic properties in the chiral soliton lattice by a minimal itinerant electron model.
Using variational calculations, we find that the period of the chiral soliton lattice can be locked at particular values dictated by the Fermi wave number, in stark contrast to spin-only models. 
We discuss this behavior caused by the spin-charge coupling as a possible mechanism for the lock-in discovered in Yb(Ni$_{1-x}$Cu$_x$)$_3$Al$_9$ [T. Matsumura {\it et al.}, J. Phys. Soc. Jpn. \textbf{86}, 124702 (2017)].
We also show that the same mechanism leads to the spontaneous formation of the chiral soliton lattice even in the absence of the magnetic field.
}

\begin{document}
\maketitle

Chiral magnets have recently attracted considerable attention, owing to their peculiar properties arising from noncollinear and noncoplanar spin textures.
In these systems, the chirality in magnetism is generated by the crystal symmetry and the spin--orbit coupling. 
A well-known interaction, which gives rise to such magnetic chirality, is the Dzyaloshinskii--Moriya (DM) interaction~\cite{Dzyaloshinskii1958, Moriya1960}.
The DM interaction, which is expressed by $\mathbf{D} \cdot \mathbf{S}_i \times \mathbf{S}_j$ ($\mathbf{D}$ is the DM vector determined by the lattice structure and $\mathbf{S}_i$ is the magnetic moment at site $i$), brings about a twist between the magnetic moments, and leads to noncollinear and noncoplanar spin textures.

\begin{figure}[hb]
\centering
\includegraphics[width=\columnwidth,clip]{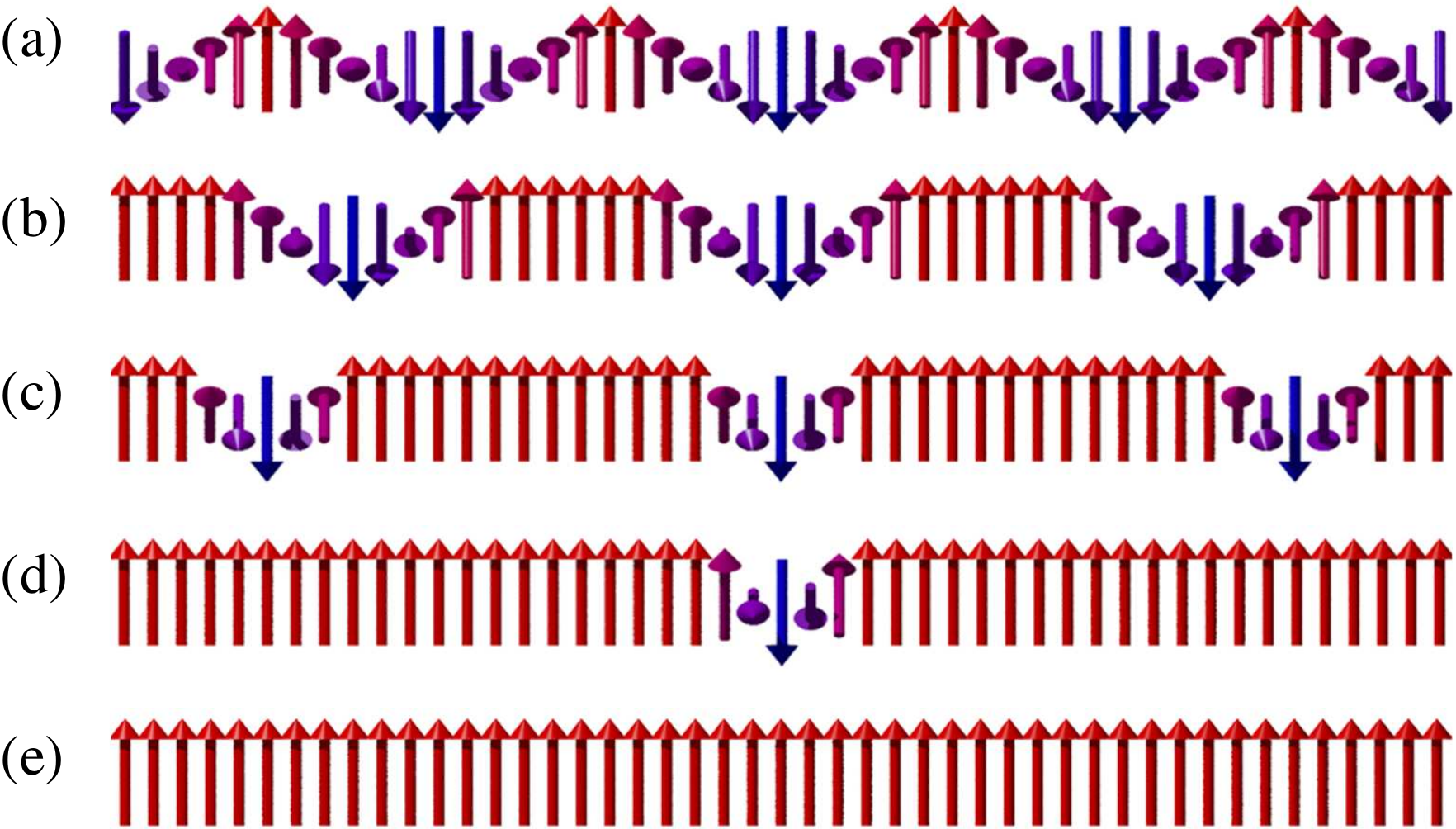}
\caption{\label{fig1} Schematic pictures of (a) a chiral helimagnetic state (CHM), (b)-(d) chiral soliton lattices (CSLs), and (e) a forced ferromagnetic state (FMM). 
The chiral axis is taken in the chain direction.
}
\end{figure}

The chiral soliton lattice (CSL) is an archetype of such peculiar spin textures. 
The CSL is a periodic array of chiral spin twists spaced by almost ferromagnetic regions, as shown in Fig.~\ref{fig1}. 
It is realized in monoaxial chiral magnets as follows.
In the absence of magnetic field, monoaxial chiral magnets exhibit a one-dimensional chiral helimagnetic state (CHM) with a uniform spin spiral structure [Fig.~\ref{fig1}(a)]. 
When a magnetic field is applied perpendicular to the helical axis, the CHM turns into a CSL, whose period increases as the magnetic field increases [Figs.~\ref{fig1}(b)-\ref{fig1}(d)], and finally relaxes into a forced ferromagnetic state (FFM) above a critical field [Fig.~\ref{fig1}(e)].
The CSL is experimentally found in chiral magnetic conductors, such as ${\rm CrNb_3S_6}$~\cite{Moriya1982, Miyadai1983, Togawa2012} and ${\rm Yb(Ni_{1-{\it x}}Cu_{{\it x}})_3Al_9}$~\cite{Matsumura2017}.

The CSL has been studied theoretically for a long time since the pioneering works by Dzyaloshinskii~\cite{Dzyaloshinskii1964, Dzyaloshinskii1965}. 
Various methods have been applied to this problem, such as a continuum approximation~\cite{Kishine2005,Kishine2010, Kishine2014, Kishine2015, Laliena2016PRB93, Laliena2016PRB94}, mean field approximation~\cite{Shinozaki2016}, and Monte Carlo simulation~\cite{Nishikawa2016}.
These studies are mostly focused on the magnetic properties of CSL on the basis of effective spin-only models while omitting itinerant electrons in the chiral magnetic conductors. 
However, the CSL exhibits intriguing behavior also in the electronic properties, such as the nonlinear negative magnetoresistance~\cite{Togawa2013}, which obviously demands further consideration including itinerant electrons. 
In addition, a peculiar lock-in of the CSL was recently discovered in Yb(Ni$_{0.94}$Cu$_{0.06}$)$_3$Al$_9$~\cite{Matsumura2017}: the period of the CSL is locked at a particular length while changing the magnetic field.
This behavior also may not be explained by the spin-only model.
The authors recently proposed a minimal model with itinerant electrons, which well reproduces the formation of CSL and the nonlinear negative magnetoresistance proportional to the soliton density~\cite{Okumura2017JPSJ}. 
It would be intriguing to further investigate the minimal model for understanding of the fundamental electronic and magnetic properties of CSL. 

In this Letter, as a complementary study to the previous Monte Carlo simulation at finite temperatures~\cite{Okumura2017JPSJ}, we investigate the ground-state properties of the minimal model, a Kondo lattice model with the DM interaction.
Performing variational calculations, we show that itinerant electrons modify the response to the magnetic field from that in the spin-only models.
In particular, we find that the period of the CSL can be locked at particular values dictated by the Fermi wave number.
The CSL with a locked period is stabilized by gap opening in the electronic structure through the scattering of electrons by chiral solitons in a finite range of magnetic field. 
We discuss the results in comparison with the lock-in observed in ${\rm Yb(Ni_{0.94}Cu_{0.06})_3Al_9}$ in detail.
We also find that the same mechanism leads to a spontaneous formation of the CSL even in the absence of the magnetic field.

Following the previous study~\cite{Okumura2017JPSJ}, we consider the ferromagnetic Kondo lattice model with the DM interaction between the localized classical spins in one dimension.
The Hamiltonian is given by
\begin{align}
H = &-t\sum_{l,\mu}(c^{\dagger}_{l\mu}c^{\;}_{l+1\mu}+\mathrm{h.c.})-J\sum_{l,\mu,\nu}c^{\dagger}_{l\mu}{\boldsymbol \sigma}_{\mu\nu}c^{\;}_{l\nu}\cdot{\mathbf S}_{l}\nonumber\\
&-{\mathbf D}\cdot\sum_{l}{\mathbf S}_{l}\times{\mathbf S}_{l+1}-h\sum_{l}S_l^{x},
\label{eq:H_el}
\end{align}
in the same notations as the previous study~\cite{Okumura2017JPSJ}. 
The model is composed of the kinetic energy of itinerant electrons with the nearest-neighbor hopping $t$, the onsite coupling between the itinerant electrons and localized classical spins with the coupling constant $J$, the DM interaction with the DM vector $\bold{D} = D\hat{z}$, where $D>0$ and $\hat{z}$ is a unit vector along the chain direction, and the Zeeman coupling of localized spins to an external magnetic field $h$ perpendicular to the chain direction, taken as $\hat{x}$.

We investigate the ground state of the model in Eq.~(\ref{eq:H_el}) by variational calculations.
Following the previous studies in a continuum approximation~\cite{Kishine2005, Kishine2010, Kishine2015}, we assume the localized spin configuration as $\bold{S}_{l}=(\cos\theta_{l},\sin\theta_{l},0)$ with
\begin{align}
\theta_{l}=\pi+2\mathrm{am}\Big{(}\frac{2K(\kappa)}{L}l\Big{)},
\label{eq:spin_config}
\end{align}
where $\mathrm{am}$ is the Jacobi amplitude function, $\kappa$ represents the elliptic modulus $(0\leq\kappa\leq1)$, $K(\kappa)$ is the complete elliptic integral of the first kind, and $L$ is the period of magnetic structure.
The spin configuration includes the CHM at $\kappa=0$, the CSL for $0<\kappa<1$, and the FFM at $\kappa=1$.  
In Eq.~(\ref{eq:spin_config}), we set a down spin ($\theta=\pi$) at the origin $l=0$ (except for the FFM); a phase shift of the spin configuration does not alter the following results qualitatively, although it leads to discretization due to the lattice effect for much smaller $L$ than studied here~\cite{footnote2}. 
For a given spin configuration, we can compute the total energy of the system per site by the exact diagonalization of the one-body Hamiltonian. 
By assuming the periodic boundary condition, the result is given by
\begin{align}
E(\kappa,L) = \frac{1}{N}\Big[\sum_{\varepsilon_{i} \leq \varepsilon_{\mathrm{F}}}\varepsilon_{i}-\sum_{l}\big\{D\sin(\theta_{l+1}-\theta_{l})+h\cos\theta_{l}\big\}\Big],
\label{eq:E_el}
\end{align}
where $\varepsilon_{i}$ is the $i$th eigenvalue and $\varepsilon_{\mathrm{F}}$ is the Fermi energy; $N$ is the number of sites, which is set at $L\times10^{4}$ in the following calculations (we take the lattice constant as the length unit). 
In the variational calculations, we optimize $E(\kappa,L)$ by varying $\kappa$ and $L$ for particular electron filling $n$ defined by $n=\frac{1}{N}\sum_{\varepsilon_{i}\leq\varepsilon_{\mathrm{F}}}$.
In the following calculations, setting the energy unit $t=1$, we fix $D=0.035$ and choose the value of $J$ for each $n$ so that $L$  becomes 10 at zero field~\cite{Okumura2017PhysicaB}.

\begin{figure}[h]
\centering
\includegraphics[angle=90,width=\columnwidth,clip]{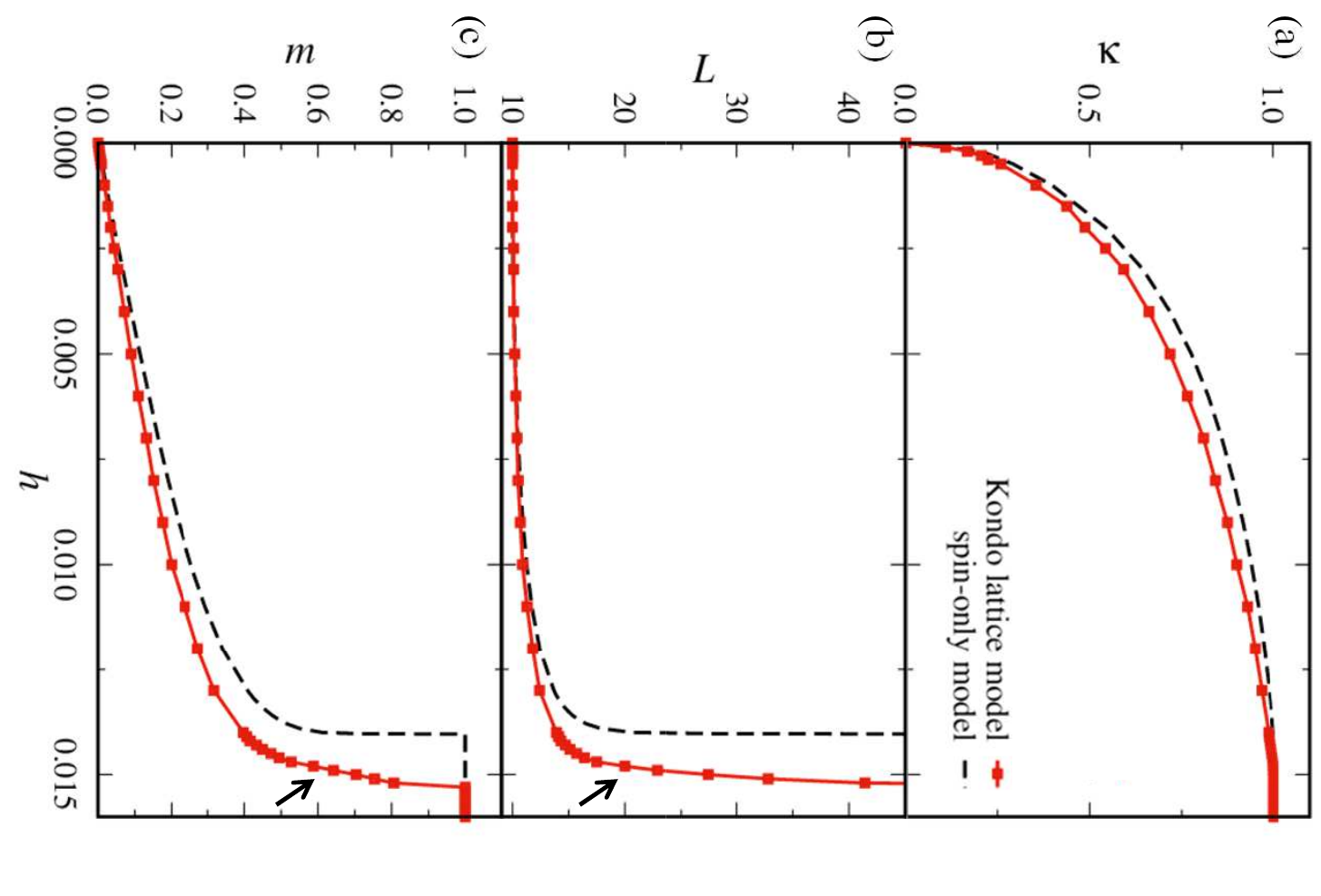}
\caption{\label{fig2} 
Magnetic field dependences of (a) the elliptic modulus $\kappa$, (b) the period of the magnetic structures {\it L}, and (c) the magnetization of the localized spins {\it m} obtained by variational calculations for the model in Eq.~(\ref{eq:H_el}) at $J=2.03$, $D=0.035$, and $n=0.5$. 
The dashed (black) line represents the result for the spin-only model for comparison~\cite{footnote1}. 
The arrows in (b) and (c) indicate the bumps discussed in the main text.
}
\end{figure}

Figure~\ref{fig2} shows the results of the variational calculation for the model in Eq.~(\ref{eq:H_el}) at quarter filling $n=0.5$ while changing the magnetic field $h$. 
Figure~\ref{fig2}(a) indicates that the optimized spin configuration changes from the CHM with $\kappa=0$ at $h=0$ to the CSL with a nonzero $\kappa$ by switching on $h$, and finally to the FFM with $\kappa=1$ at the critical field $h=h_{\mathrm{c}}\simeq 0.0153$. 
Accordingly, the period $L$ increases from the initial value $10$ and rapidly diverges while approaching $h_{\mathrm{c}}$ [Fig.~\ref{fig2}(b)]; 
the magnetization of the localized spins $m$ also increases from $0$ and rapidly saturates to $1$ as $h \to h_{\mathrm{c}}$ [Fig.~\ref{fig2}(c)].
For comparison, we show the results for the spin-only model, the ferromagnetic Heisenberg model with the DM interaction, by dashed curves in Fig.~\ref{fig2}~\cite{footnote1}.
The comparison indicates that the increase of $\kappa$, $L$, and $m$ is slower in the model with itinerant electrons than the spin-only model.
This suggests that the effective ferromagnetic interaction mediated by itinerant electrons is reduced by the magnetic field through the change in the electronic state.
We also note that the data for the itinerant electron model show bumpy behavior at $h\sim 0.015$ [indicated by arrows in Figs.~\ref{fig2}(b) and \ref{fig2}(c)]; we will return to this later.

\begin{figure}[t]
\centering
\includegraphics[angle=90,width=\columnwidth,clip]{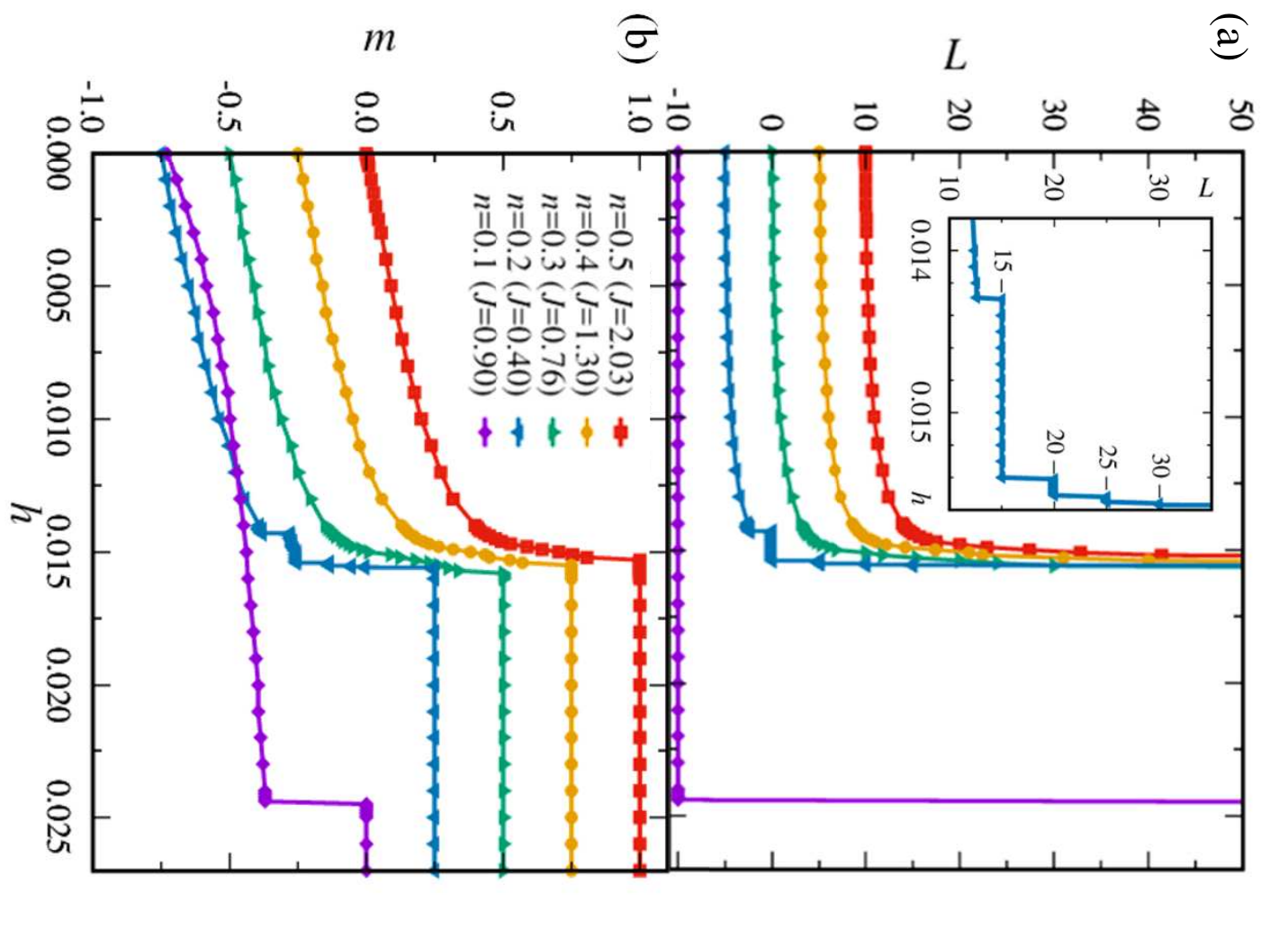}
\caption{\label{fig3} 
Magnetic field dependences of (a) {\it L} and (b) {\it m} for the model in Eq.~(\ref{eq:H_el}) at several electron fillings. 
We set $J$ for each $n$ to realize $L=10$ at $h=0$ with fixed $D=0.035$. 
For clarity, we plot the data for different $n$ with the offsets $(n-0.5)\times50$ for $L$ and $(n-0.5)\times2.5$ for $m$.
The inset of (a) presents an enlarged view of the data at $n=0.2$ around $h=0.015$. 
}
\end{figure}

Figure~\ref{fig3} shows the data of $L$ and $m$ for several electron fillings $n$.
While $n$ decreases from $0.5$ to $0.3$, the growth of $L$ and $m$ with $h$ becomes slower gradually, and the critical magnetic field $h_{\mathrm{c}}$ becomes larger.
At the same time, the bumpy behavior near $h_{\mathrm{c}}$ becomes more conspicuous for lower $n$.
For $n=0.2$, however, $L$ shows the plateaux at $L=15$, $20$, $25$, and $30$ [see also the inset of Fig.~\ref{fig3}(a)], and correspondingly, $m$ also shows plateau-like features.
We note that $L$ and $m$ change discontinuously between the plateaux. 
The data indicate that the bumpy behavior for larger $n$ is a remnant of the plateaux and the discontinuous jumps.

When further lowering $n$ to $n=0.1$, $L$ is fixed at $10$ from $h=0$ to $h_{\mathrm{c}}\simeq0.0245$.
Interestingly, $m$ is nonzero ($m\simeq 0.27$) even at $h=0$, which means a spontaneous formation of the CSL in the absence of the magnetic field (we confirmed that $\kappa$ is also nonzero). 
Thus, at this low filling, the system is in the CSL state from $h=0$ to $h_{\mathrm{c}}$ with a fixed period, while $m$ (and $\kappa$) increases gradually

\begin{figure}[!h]
\centering
\includegraphics[width=0.9\columnwidth,clip]{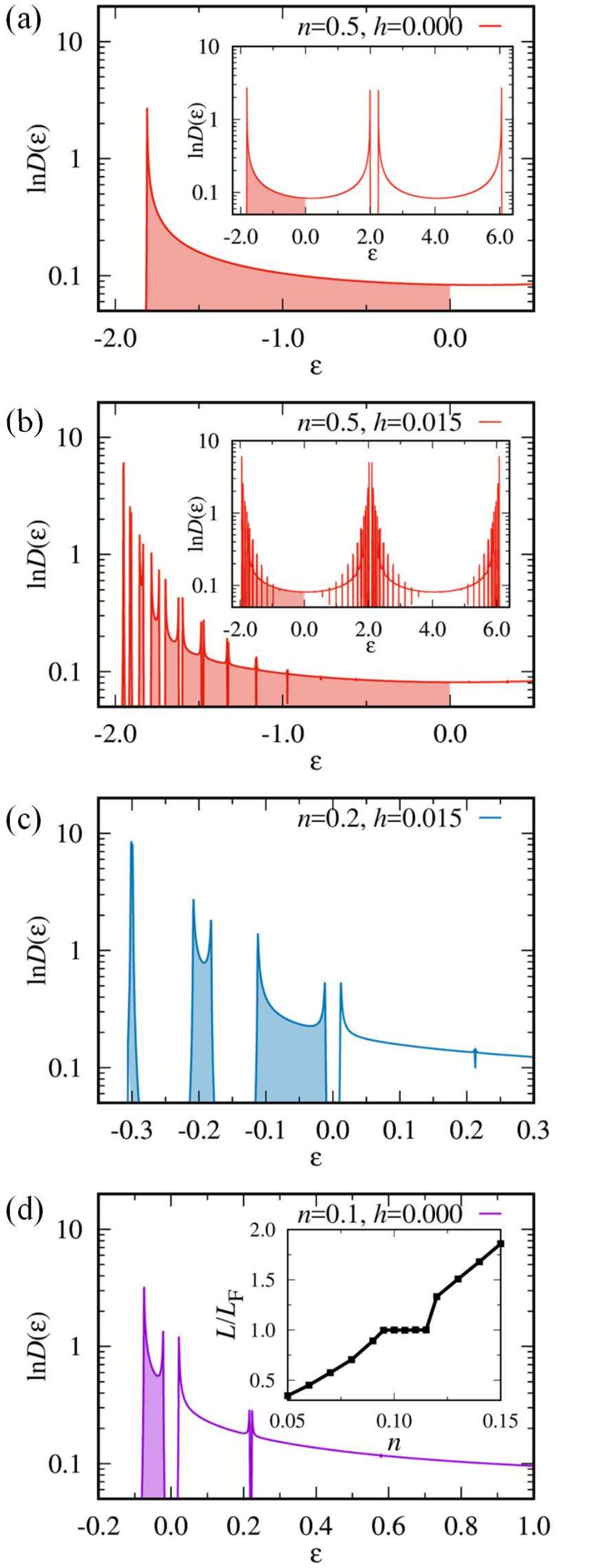}
\caption{\label{fig4} 
Density of states near the band bottom calculated for (a) the CHM at $n=0.5$ and $h=0$, (b) the CSL at $n=0.5$ and $h=0.015$, (c) the  
CSL locked at $L=15$ at $n=0.2$ and $h=0.015$, and (d) the spontaneous CSL at $n=0.1$ and $h=0$. 
The insets of (a) and (b) show the whole spectra. 
The shaded regions represent the occupied states below the Fermi energy.
The inset of (d) presents the electron filling $n$ dependence of the periodicity $L$ divided by $L_{\mathrm{F}}=\pi/k_{\rm F}$ with fixed $J=0.9$ and $D=0.035$.
}
\end{figure}

In order to clarify the origin of these peculiar behaviors appearing at low $n$, which are not seen in the spin-only model, we investigate the electronic state by calculating the density of states (DOS). 
Figure~\ref{fig4} shows the DOS for the optimized spin configurations at several different parameters.
Figure~\ref{fig4}(a) is the result for the CHM ($h=0$) at $n=0.5$, which indicates that the DOS is smooth except for the band edges. 
This DOS resembles to that for a noninteracting model because the Hamiltonian in Eq.~(\ref{eq:H_el}) with the CHM is reduced to a simple tight binding model with a renormalized hopping by a spin-dependent gauge transformation ($c_{l\mu} \to \tilde{c}_{l\mu} = c_{l\mu} e^{i \mu Q l/2}$ where $Q=2\pi/L$).
On the other hand, in the CSL at $h=0.015$, the DOS shows several energy gaps densely distributed near the band edges. 
The gaps originate from the scattering of electrons by the superlattice of chiral solitons: they open at the boundaries of the folded Brillouin zone in the CSL. 
While further increasing $h$, the number of gaps increases and the gaps come across the Fermi energy. 
This leads to the bumpy behavior near $h_{\mathrm{c}}$ in Figs.~\ref{fig2} and \ref{fig3}. 

For lower $n$ where the Fermi energy comes closer to the band bottom, the system is affected by the gap opening more conspicuously.
Figure~\ref{fig4}(c) shows the result at $n=0.2$ and $h=0.015$, where the CSL is locked at $L=15$. 
In this case, the Fermi energy is inside one of the gaps, the third-lowest one. 
Similarly, we confirmed that the CSLs locked at $L=20$, $25$, and $30$ have the Fermi energy inside the 4, 5, and 6th-lowest gap, respectively. 
Thus, the lock-in takes place when $2k_{\rm F}$ ($k_{\rm F}$ is the Fermi wave number) matches with integer multiples of the width of the folded Brillouin zone, $2\pi\ell/L$; the integer $\ell$ corresponds to the number of occupied bands [$\ell=3$ in Fig.~\ref{fig3}(c)].
Hence, for a given $n$, the lock-in condition is given by
\begin{align}
L=\frac{\ell}{n}.
\label{eq:lock-in_condition}
\end{align}

Figure~\ref{fig4}(d) shows the DOS at $n=0.1$ and $h=0$, where the CSL is formed spontaneously.
In this state, the Fermi energy is inside the lowest gap already at zero field, as Eq.~(\ref{eq:lock-in_condition}) is satisfied with $\ell=1$. 
Hence, the spontaneous formation of the CSL is understood by the common mechanism to the lock-in found for higher $n$ at nonzero field. 
While changing $n$ at $h=0$, it occurs when $n$ is in the range where $2k_{\rm F}=2\pi n$ matches with $2\pi/L$. 
The inset of Fig.~\ref{fig4}(d) shows such a range of $n$ with $L/L_{\rm F} = L/(\pi/k_{\rm F}) =1$ for fixed $J=0.9$. 

Let us discuss our results in comparison with experiments. 
In Yb(Ni$_{0.94}$Cu$_{0.06}$)$_3$Al$_9$, a lock-in of the CSL was found at the period of eight Yb moments along the chiral axis. 
The lock-in was observed only at eight, although the period changes from $\sim 6.8$ to $\sim 8.4$ in the measured range of magnetic field. 
The lock-in with the lack of the period seven is difficult to explain by the lattice discretization, as it predicts the lock-in at every integer~\cite{footnote1}. 
Meanwhile, our mechanism predicts that the lock-in occurs only at integer multiples of a particular value determined by the electron filling.
Hence, the origin of the lock-in found in Yb(Ni$_{0.94}$Cu$_{0.06}$)$_3$Al$_9$ would be ascribed to our mechanism from the coupling to itinerant electrons.
If this is the case, $k_{\mathrm F}$ along the $c$-axis satisfies $2k_{\mathrm F}=2\pi/L=3\pi/(4c)$ ($c$ is the lattice constant along the chiral axis), which will be confirmed by the measurement of the Fermi surface. 
Our scenario would also be tested by the electrical transport measurement, as it predicts the high-resistive state during the lock-in. 
We note that our analysis is for the simplified one-dimensional model, but the similar mechanism could work by partial nesting of the Fermi surface in realistic three-dimensional systems. 

Thus far, a spontaneous formation of the CSL has not been found in experiments.
Our calculations suggest that lower electron filling might increase a chance to observe the spontaneous CSL.
In this sense, ${\rm Yb(Ni_{1-{\it x}}Cu_{{\it x}})_3Al_9}$ can be a promising compound, since the electron filling can be tuned by Cu doping~\cite{Matsumura2017}.

In summary, we have investigated the effect of itinerant electrons on the formation of CSL by variational calculations, in comparison with the spin-only model.
We clarified how the coupling to itinerant electrons modifies the development of the CSL in an applied magnetic field. 
In particular, we found that the period of CSLs can be locked at a particular set of values. 
We elucidated that the lock-in is explained by gap opening in the electronic states due to the scattering of electrons by chiral solitons. 
We discussed the finding as a possible mechanism for the lock-in observed in Yb(Ni$_{1-x}$Cu$_x$)$_3$Al$_9$. 
We also found that the same mechanism predicts the spontaneous formation of the CSL even in the absence of the magnetic field. 
It would be interesting to clarify how the lock-in develops as a function of temperature in a realistic three-dimensional system, which will affect both electronic and magnetic properties. 
This problem is left for future study.

\begin{acknowledgments}
The authors thank J.~Kishine, K.~Inoue, S.~Ohara, and Y.~Togawa for fruitful discussions.
This research was supported by KAKENHI (No.~15K05176).
This work was also supported by the Chirality Research Center in Hiroshima University and JSPS Core-to-Core Program,
Advanced Research Networks.
Part of the computation in this work was carried out at the Supercomputer Center, Institute for Solid State Physics, the University of Tokyo.
\end{acknowledgments}


\begin{thebibliography}{9}
\bibitem{footnote1}The discretization appears at every integer of $L$, which is distinguished from the lock-in in the present study.
\end{thebibliography}
\end{document}